\documentclass{moriond}

\def\Journal#1#2#3#4{{#1} {\bf #2}, #3 (#4)}

\def\PLB{{\em Phys. Lett.}  B}
\def\PRL{\em Phys. Rev. Lett.}
\def\PRD{{\em Phys. Rev.} D}

\def\fbinv{{\rm fb}^{-1}}
\def\pbinv{{\rm pb}^{-1}}
\def\pt{p_{\rm T}}
\def\ttbar{{\rm t\bar{t}}}
\def\mtop{m_{\rm t}}
\def\as{\alpha_{\rm S}}

\newcommand{\Photo}{\includegraphics[height=35mm]{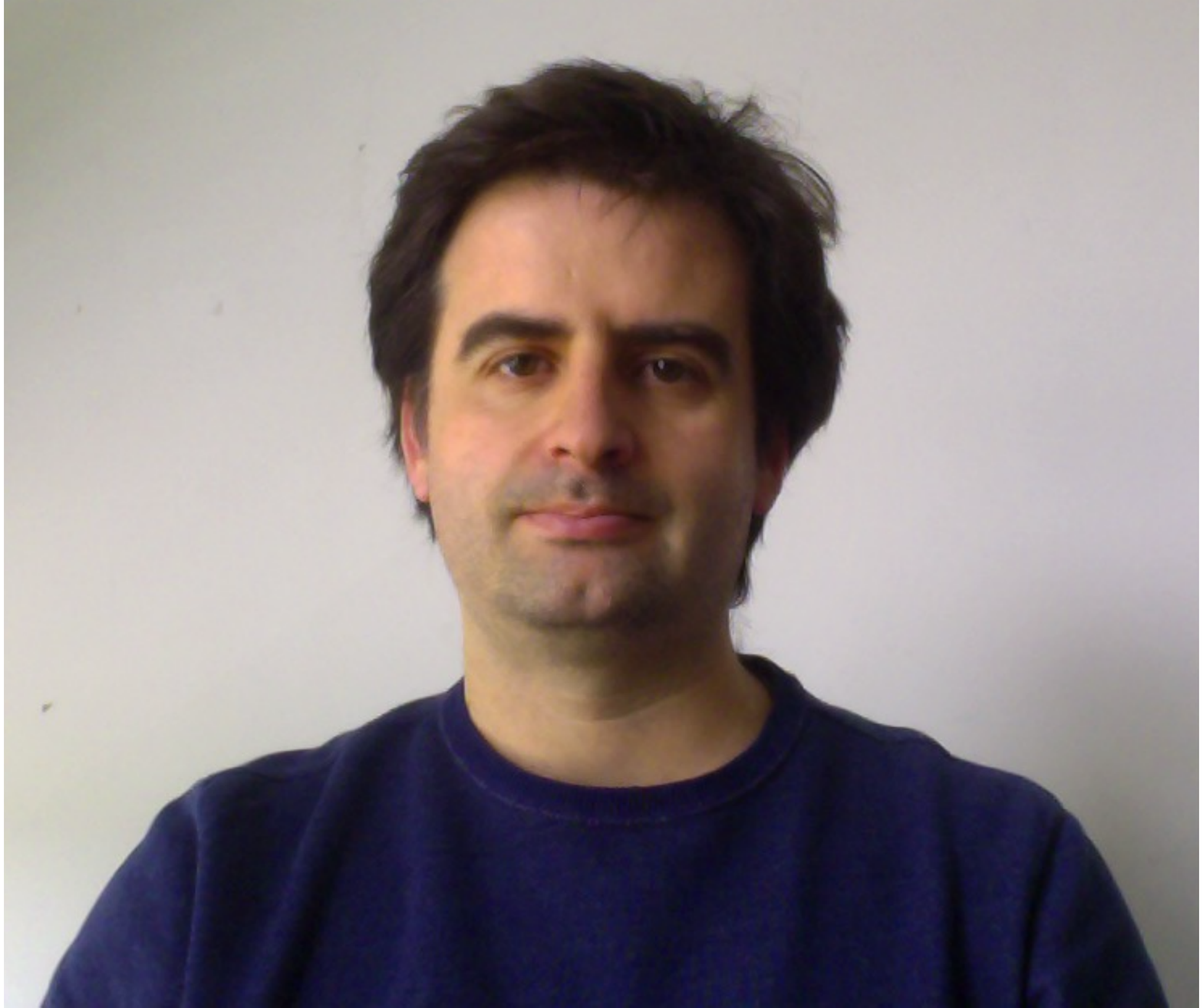}}

\begin{document}

\vspace*{4cm}
\title{Top quark production at the LHC}

\author{P. Ferreira da Silva (CERN),\\\em on behalf of the ATLAS and CMS Collaborations}
\address{~}

\maketitle\abstracts{
Twenty years past its discovery, the top quark continues attracting
great interest as experiments keep unveiling its properties. An overview of
the latest measurements in the domain of top quark production,
performed by the ATLAS and CMS experiments at the CERN LHC, is given. 
The latest measurements of top quark production rates via strong and electroweak
processes are reported and compared to different perturbative QCD
predictions. Fundamental properties, such as the mass or the couplings of the
top quark, as well as re-interpretations seeking for beyond the
standard model contributions in the top quark sector, are extracted
from these measurements. In each case an attempt to highlight
the first results and main prospects for the on-going Run 2 of the LHC is
made.}


%
%
%
\section{Introduction}
\label{sec:intro}

\begin{figure}
\hfill
\begin{minipage}{0.545\linewidth}
\centerline{\includegraphics[width=1.0\linewidth]{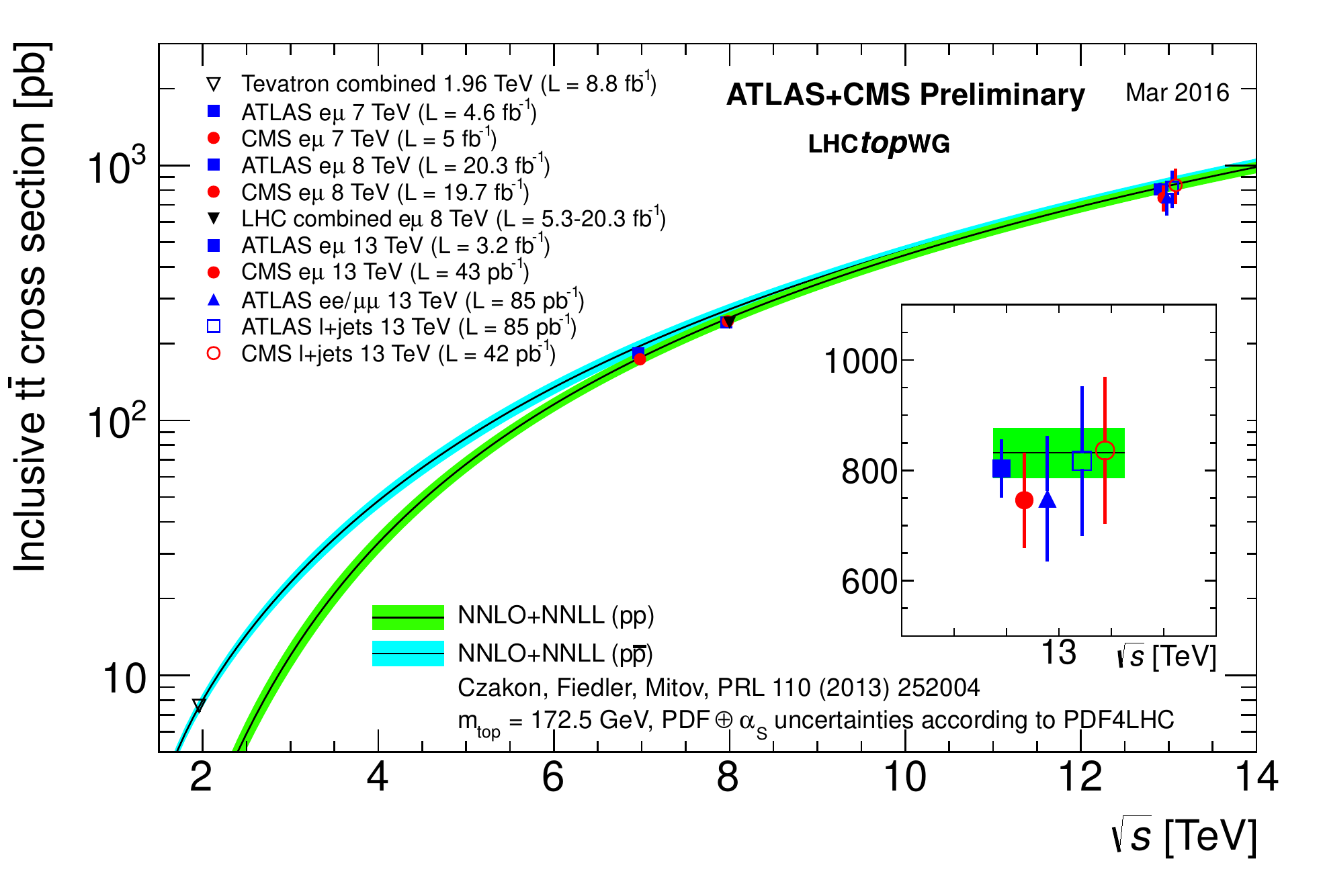}}
\end{minipage}
\hfill
\begin{minipage}{0.445\linewidth}
\centerline{\includegraphics[width=1.0\linewidth]{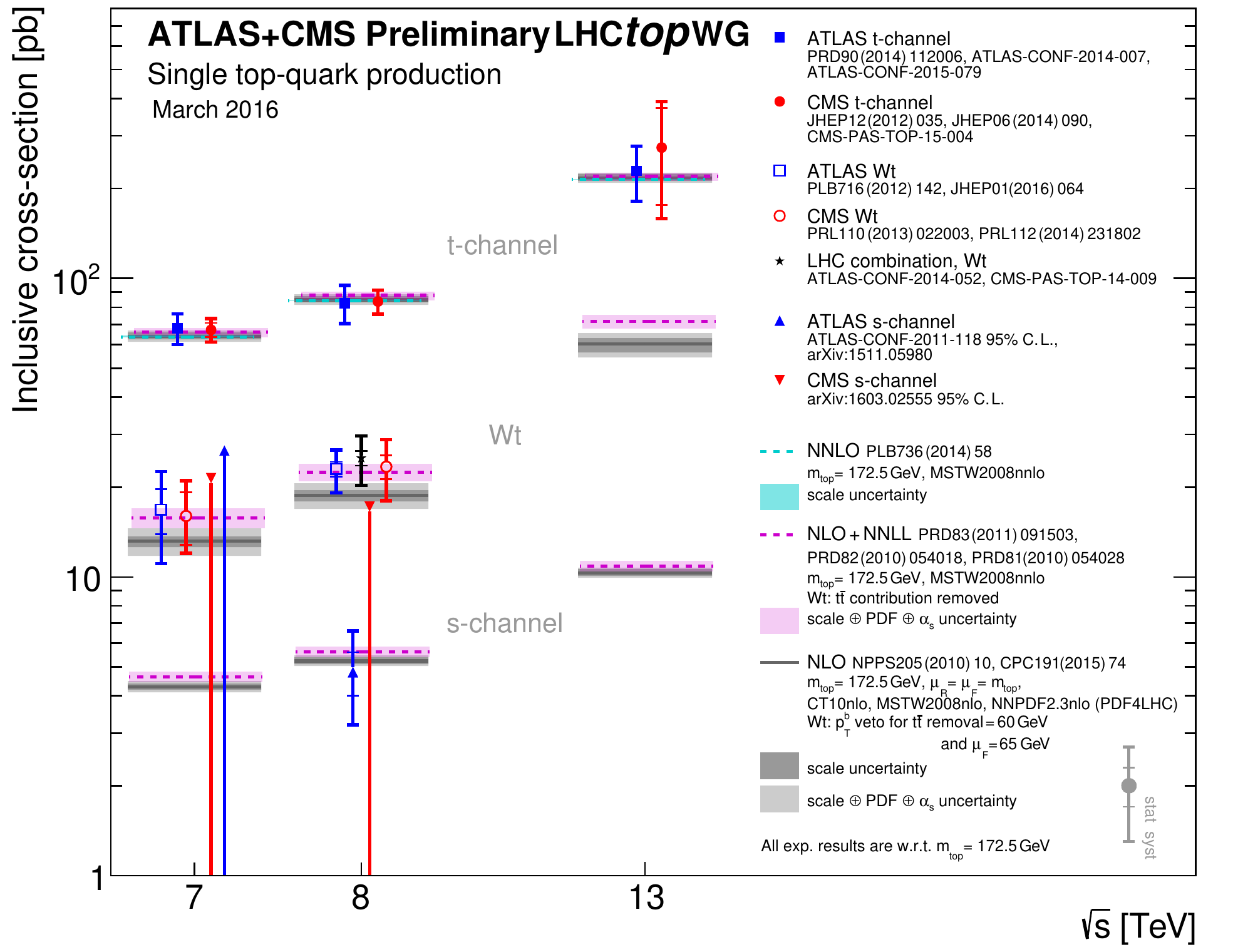}}
\end{minipage}
\caption[]{Left: evolution of the $\ttbar$ cross section as function
  of $\sqrt{s}$ at NNLO+NNLL compared to different measurements. The
  inset highlights the most recent 13 TeV measurements. Right: summary
  of single top production cross section measurements in the different
  production modes at different $\sqrt{s}$.}
\label{fig1}
\end{figure}

Early measurements of top quark production in proton-proton collisions
at $\sqrt{s}=$13 TeV  are showing
overall good agreement with the standard model predictions.
These results open the door to a new era where the properties of the
top quark will be probed to a new level in precision, profiting from a
high integrated luminosity dataset expected to be collected after Run 2
of the LHC ($>100~\fbinv$).
In this writeup the latest results regarding the study of strong and electroweak
top quark production modes
are reviewed.
The available results for the inclusive cross section measurements,
at the time the talk has been given, are summarised in Figure~\ref{fig1}.
A selection of these results is made, with the intention
of highlighting and discussing the main uncertainties.
These early results can
already provide information on the main experimental and theory
uncertainties which need to be improved in order to achieve the
necessary precision for top quark physics at the end of Run 2.

\section{Strong production of top-antitop quark pairs}
\label{sec:ttprod}

At hadron colliders top quarks are predominantly produced in pairs
($\ttbar$)  through strong interactions. 
The state of the art calculations are able to predict the rates at
which $\ttbar$ pairs are produced at next-to-next-to-leading order,
including soft-gluon resummation to next-to-next-to-leading-log order
(NNLO+NNLL)
with an uncertainty of $\approx$5\%~\cite{top++}.
With respect to the first run of the LHC at 7 and 8 TeV center of mass
energy, an increase by a factor of $\approx$3.3 in rate is expected,
for the 13 TeV pp collisions.

The $\ttbar$ production cross section is expected to decrease
approximately with the fourth power of the top quark mass ($\mtop$),
a characteristic which can be explored experimentally to measure the latter.
If measured differentially, the production cross section may be
sensitive to the top quark width, and electroweak corrections which
depend, amongst other factors, on the top quark Yukawa coupling to the Higgs boson.
The typical timescale of the event is dictated by a fast electroweak
decay (${\rm t\rightarrow Wb}$) which occurs within 10$^{-25}$s.
This is shorter, by an order of magnitude, than the typical
hadronization timescale ($\sim 1/\Lambda_{\rm QCD}$) and leads
to the preservation of the properties of a bare quark in the final
states.

\begin{equation}
\overbrace{\frac{1}{\mtop}}^{\stackrel{production}{10^{-27} \rm s}} ~<~
\overbrace{\frac{1}{\Gamma_{\rm t}}}^{\stackrel{lifetime}{10^{-25}\rm s}}~<~
\overbrace{\frac{1}{\Lambda_{\rm QCD}}}^{\stackrel{hadronization}{10^{-24}\rm s}}~<~
\overbrace{\frac{\mtop}{\Lambda_{\rm QCD}^2}}^{\stackrel{spin-flip}{10^{-21}\rm s}}
\end{equation}

The final states contain b-jets and other jets or leptons from the W
decays and are usually classified to the lepton multiplicity as: fully
hadronic (46\%), $\ell$+jets (45\%) or dilepton final states (9\%).

\subsection{Inclusive cross section measurements}
\label{subsec:incxsec}

During the Run 1 of the LHC, the experiments have explored many final states to
measure the $\ttbar$ cross section, and all have been
found to be in agreement with each other and with the theory.
The dilepton channel was found to lead in precision,
owing to its purity (typically $>$90\%)
and loose selection, for which only a small extrapolation factor is
required. The latter proves also to be one of the most relevant
figures of merit for the extraction of the pole mass, as the final
acceptance is expected to have minimal dependency on the top mass
itself, owing to loose kinematics requirements.

Early analysis at 13 TeV have promptly established $\ttbar$ production
with a rate close to that which is predicted at NNLO+NNLL
(see Fig.~\ref{fig1}, left).
At 13 TeV tops are now probed in a region which is dominated by
gluon-gluon fusion ($\approx$86\% of the production), 
making $\ttbar$ production a standard model (SM)
candle for gluon luminosities.
An early CMS measurement~\cite{Khachatryan:2015uqb}, using $42~\pbinv$ of
data, and electron-muon final states,
can be used to exemplify how experimental uncertainties dominate at
restart of operating the experiments.
In this analysis the leptons are required to have $\pt>$30 GeV and $|\eta|<$2.5.
The main backgrounds, consisting of
Drell-Yan (DY), dibosons (WW, WZ, ZZ) and top+W (tW) associated production,
become negligible after requiring two jets ($\pt>30 GeV$, $|\eta|<2.5$) in the event. 
The uncertainty attained in this early measurements (11\%) is
dominated by statistics (7.8\%) the luminosity (4.8\%) and efficiency
estimation-related (5.8\%) uncertainties. These are expected to
improve in the long run, 
in particular the uncertainty on the trigger and selection
efficiencies which are derived from control regions such as
$Z\rightarrow\ell\ell$ events and are therefore expected to benefit from higher statistics.

With the full integrated luminosity of the 2015 dataset the
measurements have started to constrain in-situ some of the systematics by adopting
fitting strategies to the yields observed in different event
categories.
The latest result from ATLAS~\cite{ATLAS-CONF-2016-005}
makes use of electron-muon final states and 
counts the number of events with one or two jets identified as b-jets (b-tagged).
The events, counted in these categories, can be related amongst each
other by assuming $V_{\rm tb}=1$ and by writing the probability of 
selecting and $b$-tagging jets from top decays, using a binomial expansion
of the efficiency of the ``finding'' these jets. 
As an example the number of $\ttbar$ events with 2 b-tagged jets is expected to
be: $N_{2}(\ttbar)=\mathcal{L}\sigma_{\rm \ttbar}\varepsilon_{e\mu}\varepsilon_{b}^2C_b$
where $\mathcal{L}$ is the integrated luminosity, $\sigma_{\rm \ttbar}$
the cross section to be extracted, $\varepsilon_{e\mu}$ the product of the acceptance
and efficiency for selecting $e\mu$ final states, $\varepsilon_{b}$
the b-finding efficiency which is determined in-situ and $C_b$ a
residual correlation factor for the efficiency of finding the two b
jets in a $\ttbar$ event. Using $b$-tagging in the analysis yields a
purer selection, while reducing the number of jets required yields
higher efficiency. The fit result has a 6.7\% total
uncertainty and it is fully consistent with the SM prediction.
In this case the uncertainty is still dominated by the luminosity determination (5.5\%),
but the next-to-leading uncertainty is due to the extrapolation to the
full phase space and it is therefore of theoretical nature.
This uncertainty is estimated
by varying the signal model (2.9\%) and includes 
a change of the hadronizer used in the simulations (\textsc{PYTHIA} versus
\textsc{HERWIG++}), a change of the NLO matrix element generator
(\textsc{POWHEG} versus \textsc{MG5\_aMC@NLO}),
and the choice of the QCD scales. 
Signal model related uncertainties can be partially mitigated by performing
fiducial measurements. In this case this source of uncertainty
decreases to 2.0\%.
However in the long run these uncertainties need better understanding,
e.g. by performing dedicated measurements of the
underlying event and jet activity, as it will ultimately be a
limiting factor once the assessment of the integrated luminosity
improves. A recent example of such measurements at 13 TeV, can be found in~\cite{CMS-PAS-TOP-15-017}.
Further  partial cancellation of some of the experimental systematics can be achieved by
computing the ratio of cross sections as, e.g. to the production of Z
bosons as performed recently by ATLAS~\cite{ATLAS-CONF-2015-049}, using
the same fit technique to extract the $\ttbar$ cross section in the
same-flavour final states (ee and $\mu\mu$). 
The ratio $\sigma(\ttbar)/\sigma({\rm Z})$ is determined to be
$R_{\ttbar/Z}=0.445\pm0.027({\rm stat})\pm0.028({\rm syst})$,
in good agreement with the SM prediction obtained with the CT10 PDF set:
$R_{\ttbar/Z}=0.427^{+0.022}_{-0.013}({\rm PDF})^{+0.012}_{-0.016}({\rm scales})^{+0.005}_{-0.004}(\alpha_{\rm S})$.
It is relevant to notice that the measurement is mainly affected, again, by the uncertainty in the
signal model ($\approx$5\%).
The result can be used to probe the gluon content predicted by the
PDFs: fair agreement is found for CT10, NNPDF3.0, MMHT while some
tension with ABM12 has been reported~\cite{ATLAS-CONF-2015-049}.

With more data it is expected that the inclusive cross section analysis will be able to explore
finer details, including differential distributions in the fits. A
recent example of how these analysis may evolve can be considered
from a CMS result using 20 fb$^{-1}$ at 8 TeV~\cite{Khachatryan:2016mqs}.
The analysis makes use of the extra jet
activity to constraint the signal modeling uncertainties in the
visible phase space. A combined fit to the distribution of the transverse
momentum of the extra jets reconstructed in the events,
categorised according to the total number of jets and $b$-tags in the event,
leads to a reduced signal modeling uncertainty (1.1\%) and to
to a significant improvement of the final
uncertainty which is observed to be 3.8\%.
Further optimisation of these techniques, along with the ability to
maintain low $\pt$ single lepton and (or) dilepton trigger thresholds
in Run 2 is expected to lead to very precise measurements of the $\ttbar$
production cross section.

\subsection{Pole mass measurements from production rates}
\label{subsec:polemass}

The more precise and close
to the full phase space the analysis are, the more adequate they are
to extrapolate fundamental parameters such as $\mtop$ or $\as$.
While performing these extrapolations it is crucial to ensure that the acceptance
has minimal sensitivity on the parameters being measured. In the
last analysis referred in the previous section,
the dependency attained is at the level of 0.4\% per
GeV~\cite{Khachatryan:2016mqs}.
In addition, when comparing the observed cross section to the theory prediction
computed at fixed $\sqrt{s}$, the uncertainty in the nominal beam
energy at the LHC 
needs to be taken into account as it is known up to $\approx$1.7\%~\cite{CERN-ATS-2013-040}.
This reflects in an additional intrinsic limitation which is assigned
to the reference cross section to be used for the extraction of the
pole mass. 
Finally the theory prediction is also bound to the PDF set which is
used. In the long run, and with improved PDF sets which incorporate
LHC data in their fits, special care must be taken in order not to
incur in a circular dependency between the parameter extracted and the
``tuned'' prediction for the composition of the proton.

An uncertainty of 1\% in the determination of the pole mass
has been attained after Run 1~\cite{Khachatryan:2016mqs}.
Optimistically, if the experimental
uncertainties can be reduced to a total of 2\% (luminosity included),
and the acceptance dependency can be kept to a value $<$1\%,
one can expect to attain 0.5\% uncertainty in the determination of
$\mtop$ from the measurement of $\sigma(\ttbar)$ in Run 2 leading to a 
more competitive result to be compared with the direct $\mtop$
measurements. 
The direct $\mtop$ measurements have been discussed in detail during this conference by
B. Stieger~\cite{stiegerb}.

\subsection{tW single resonant production and its interplay with $\ttbar$}
\label{subsec:tw}

Although it contains a single top quark in the final state, the tW
process may be generated by similar initial states and yields similar
final states as a quark-gluon induced $\ttbar$ event.
As such, the associated tW production can only be factorised with
respect to $\ttbar$ to a certain extent, up
to next-to-leading-order (NLO).
At higher orders this process must be encompassed, together
with $\ttbar$, as part of inclusive WbWb production.
Its main feature is a W boson recoiling against the top quark. 
The process has been measured by both collaborations in Run 1, 
in agreement with the SM prediction~\cite{Chatrchyan:2014tua,Aad:2015eto,CMS-PAS-TOP-14-009}.
Separation with respect to $\ttbar$ is achieved, in both cases, by
resorting to multivariate analysis which explore the difference in
kinematics of the signal.
The latest measurement, from ATLAS~\cite{Aad:2015eto}, gives a step towards understanding the
single and double resonant production of top quarks as a whole and measures the combined $\ttbar+{\rm tW}$
production in a fiducial phase space.
Events with two leptons ($\pt>25$ GeV and $|\eta|<2.5$) and
with significant missing transverse energy 
($>$40~GeV for ee/$\mu\mu$ final states, $>$20 or 50~GeV
depending on the mass of the e$\mu$ final state),
are selected.
The pseudo-rapidity of the system formed by all the jets and leptons
is required to be compatible with central production ($|\eta^{\rm sys}|<2.5$) 
and the events are categorized according to the number of jets
($\pt>$20~GeV and $|\eta|<2.5$ and
number of b-tagged jets.
An uncertainty of 8.5\% is achieved, and it is limited by experimental
uncertainties such as jet energy scale, resolution (5.2\%) and
b-tagging (2.3\%). Theory uncertainties in the modeling of tW and
$\ttbar$ contribute significantly up to 4.5\% in the fiducial cross section measurement.

\subsection{Differential cross section measurements}
\label{subsec:diffxsec}

Once the inclusive production is well established,
the large statistics can be used to explore differentially the
kinematics and characteristics of $\ttbar$ events. 
The main challenges in measuring differential $\ttbar$ cross sections are related
to the multiplicity and plethora of final states which yield several
physics objects to be associated to each top quark decay.
The algorithms used to perform this reconstruction also need to cope
with a wide range of kinematics and the large amount of extra
radiation which may be produced in association with a top quark pair.
Kinematics fitting techniques, imposing the W masses, the equality of
the top quark and anti-quark masses and that the missing transverse
energy is equal to the sum of the $\pt$ of the neutrinos produced
after the W boson decays, are often used. In case several solutions
are found for an event minimizing the $\chi^2$ of the fit,
or the reconstructed invariant mass of
the system are often used as a criteria to pick a solution, the latter
being mostly driven by steeply falling PDFs. 
A likelihood-based choice, testing the compatibility of the
reconstructed kinematics with the ones expected for correct
assignments of the final state objects to the top quark decays, can
also be used, to maximize the proability of correctly reconstructing
the kinematics to be probed~\cite{CMS-PAS-TOP-16-008}.
Nevertheless, one should notice that the ambiguity in the solutions does not
necessarily need to be lifted by a specific criteria like the ones
mentioned above. A concrete example comes from top
mass measurements employing ideogram-like
techniques~\cite{Khachatryan:2015hba}, 
where different solutions
may be accomodated by weighting according to its probability of being
correct, wrong or missing partially or completely the decay products
of a top quark. This alternative is however still to be explored for
differential cross section measurements.

The reconstructed top quark kinematics has  to be compared to some
theory prediction. Comparisons to fixed-order calculations are done by correcting the reconstructed level top quarks
to the so-called parton level definition of the top which assumes intrinsically on-shell top
quarks. On the other hand, comparisons to different MC simulations, or
fixed-order-calculations carried up to the final state particles, can
be made by correcting the reconstructed kinematics to the so-called
particle level which encapsulates the possible differences in the
modeling of the decay, hadronization and fragmentation from a top
quark to its final state products. The latter is therefore, expected
to be less prone to hadronization or parton shower model-related
uncertainties.

\begin{figure}
\hfill
\begin{minipage}{0.47\linewidth}
\centerline{\includegraphics[width=1.0\linewidth]{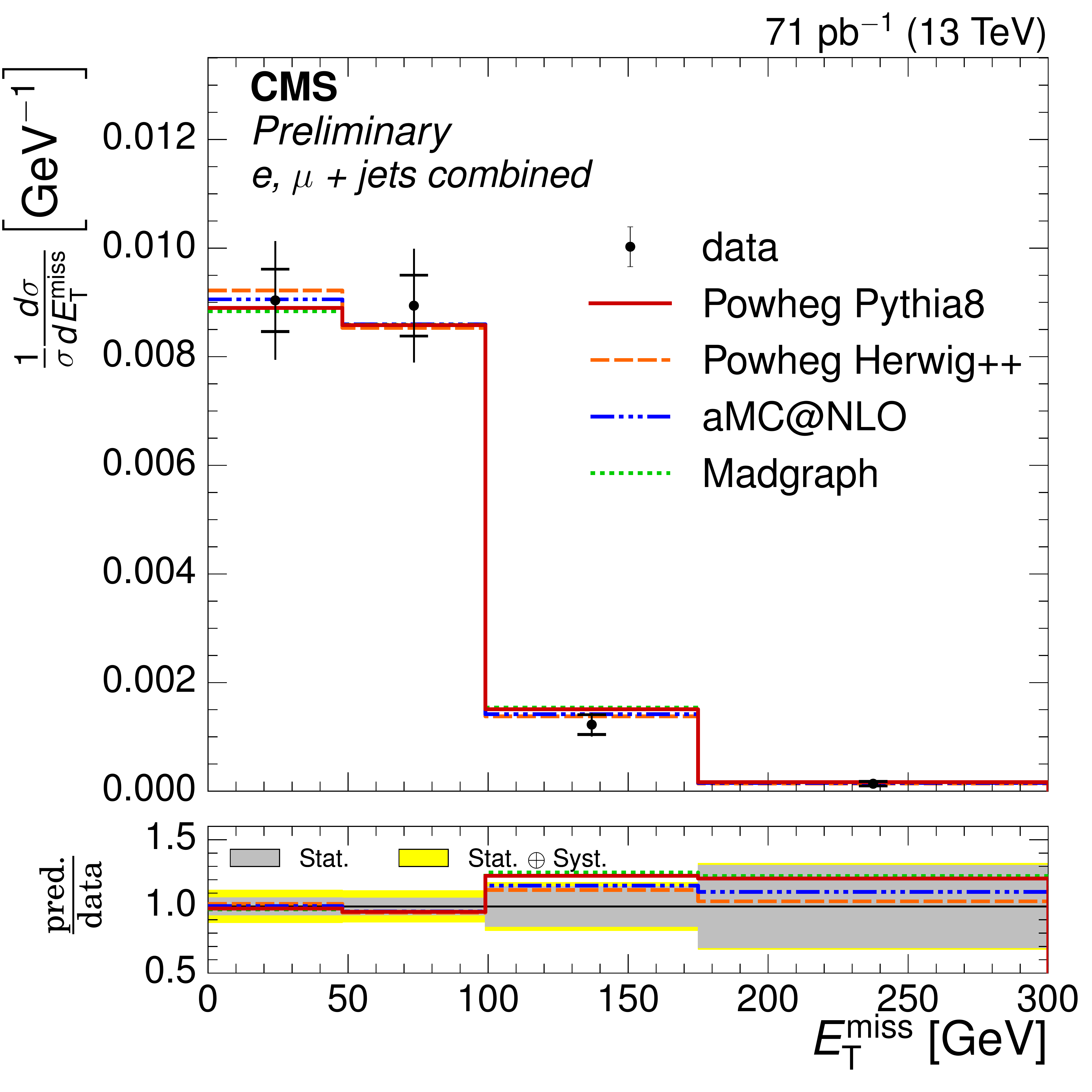}}
\end{minipage}
\hfill
\begin{minipage}{0.51\linewidth}
\centerline{\includegraphics[width=1.0\linewidth]{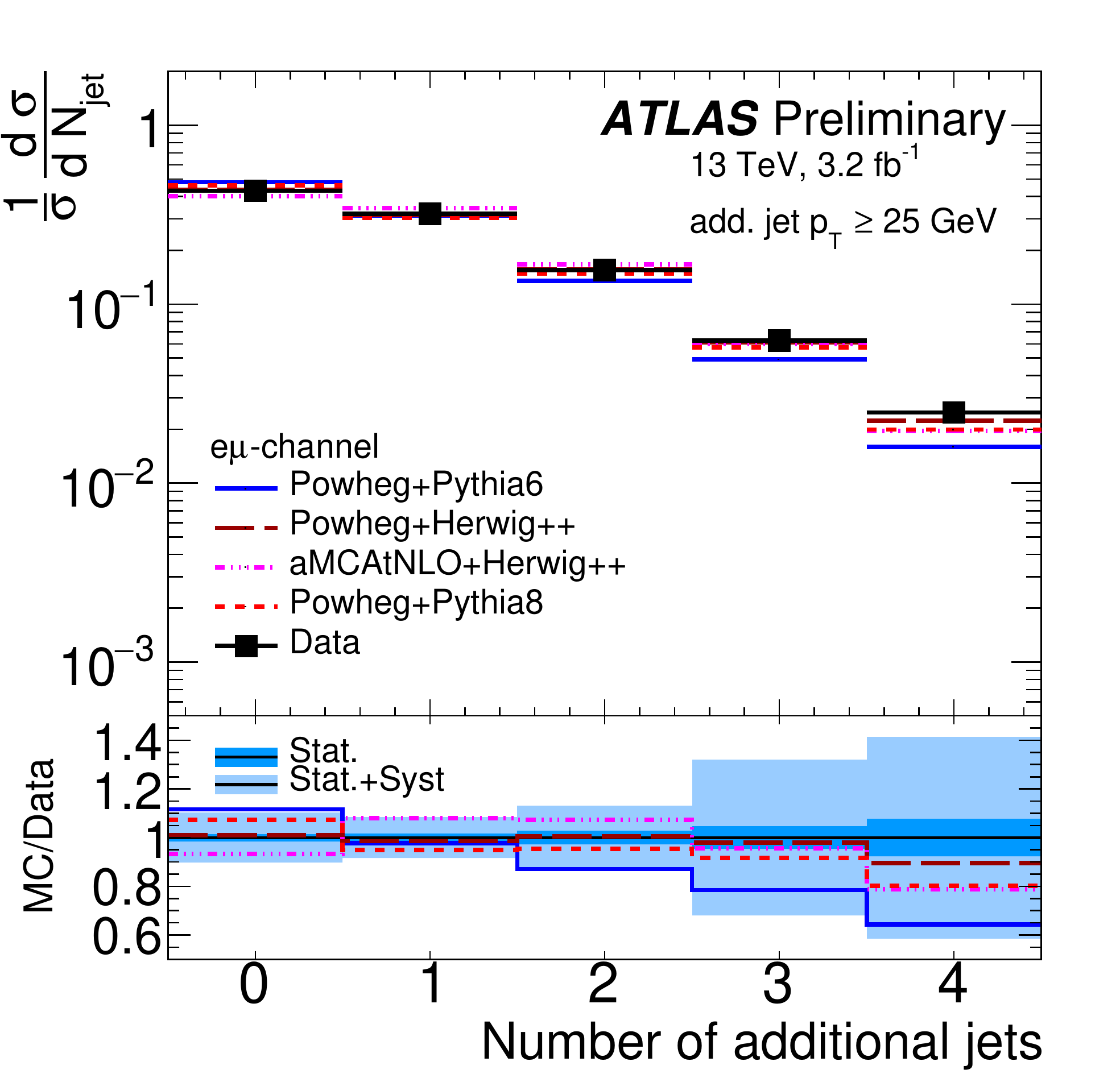}}
\end{minipage}
\caption[]{
  Left: Normalised differential cross section with respect to the
  missing transverse energy for different event generators compared to
  13 TeV data~\cite{CMS-PAS-TOP-15-013} in the $\ell$+jets final state. The inner
  (outer) bars denote the statistical (total) uncertainty.
  Right: jet multiplicity distribution for additional jets
  with $\pt>$25 GeV in the $e\mu$ channel~\cite{ATLAS-CONF-2015-065}. The statistical (total)
  uncertainties is represented by a dark (light) blue band and data
  are compared to the predictions of different parton shower
  interfaced to different matrix-element generators.
  
}
\label{fig2}
\end{figure}

An example at particle level, which does not need to reconstruct fully
the top quark kinematics, is the measurement of global event
variables such  as the missing transverse energy, the scalar sum of
the $\pt$ of the jets or the number of jets~\cite{CMS-PAS-TOP-15-013}. 
This early analysis has used the first 71 $\pbinv$ of 13 TeV data
and it's therefore limited by statistics.
The results are found to be compatible with most of the MC setups used
in Run 2.
An example is given in Figure~\ref{fig2} (left) for the missing
transverse energy distribution measured in $\ttbar$ events.
Making use of the full 2015 dataset the experiments have
started to observe~\cite{ATLAS-CONF-2015-065,CMS-PAS-TOP-16-011}
some shortcomings of the current MC setups e.g. for the modeling of
the extra jet multiplicities, in particular in regions where the
parton shower predictions are expected to dominate. Both experiments
have compared to NLO matrix element generators (including up to 1 or 2
extra partons) matched to the \textsc{PYTHIA 8} or \textsc{HERWIG++}
parton-shower generators. From the results obtained, a $>$40\%
difference is observed with respect to the nominal prediction for the number of
events with $\geq$4 additional jets with $\pt>$25-30 GeV, one can conclude
that further tuning of the parton-shower generators is needed in
the long run. Figure~\ref{fig2} (right), exemplifies one of these measurements.
A better description of $\ttbar$+jets is particularly important towards a better
modeling of $\ttbar$ as a background for $\ttbar$H or $\ttbar\ttbar$ final states.

\begin{figure}
\hfill
\begin{minipage}{0.49\linewidth}
\centerline{\includegraphics[width=1.0\linewidth]{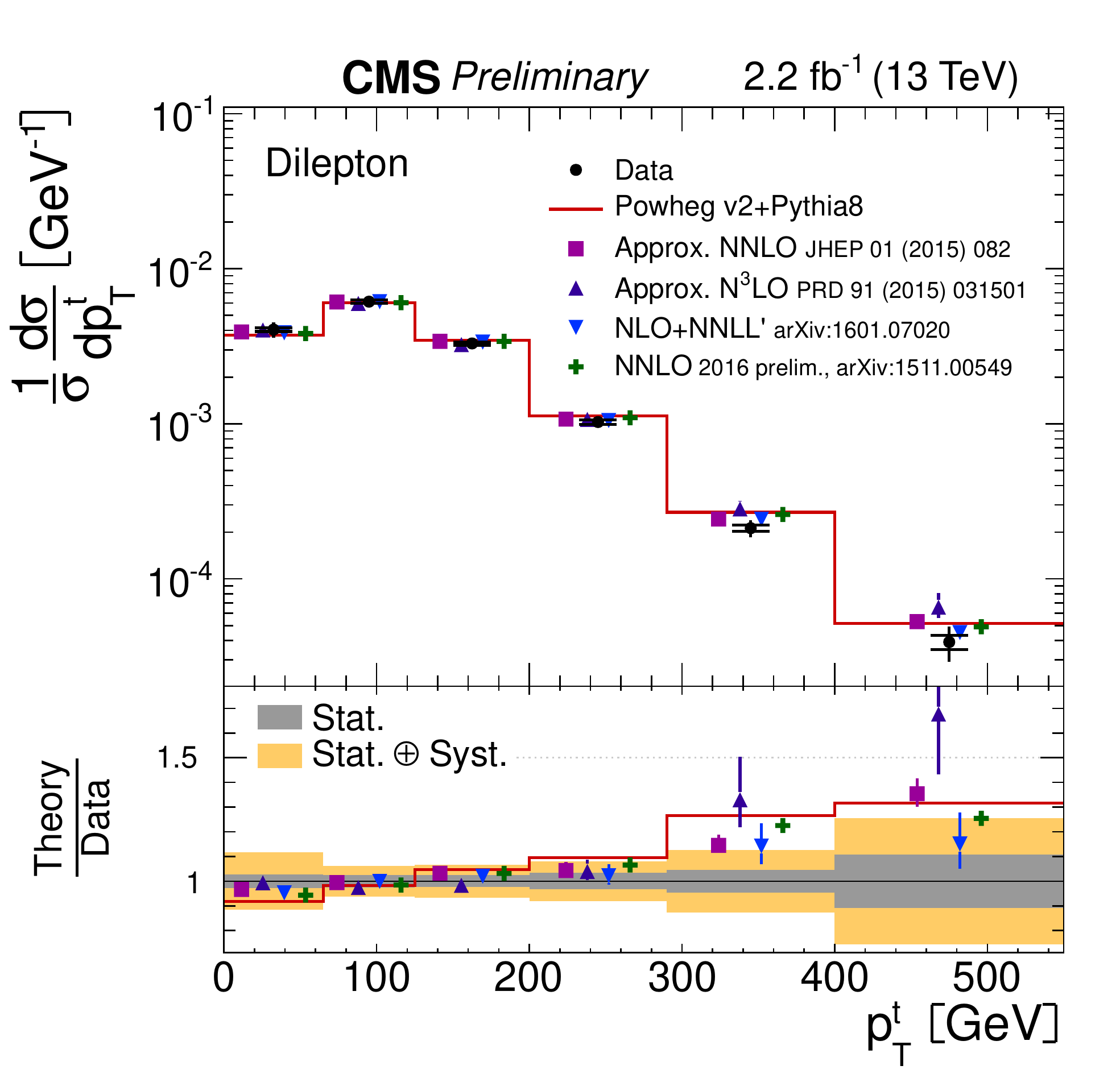}}
\end{minipage}
\hfill
\begin{minipage}{0.49\linewidth}
\centerline{\includegraphics[width=1.0\linewidth]{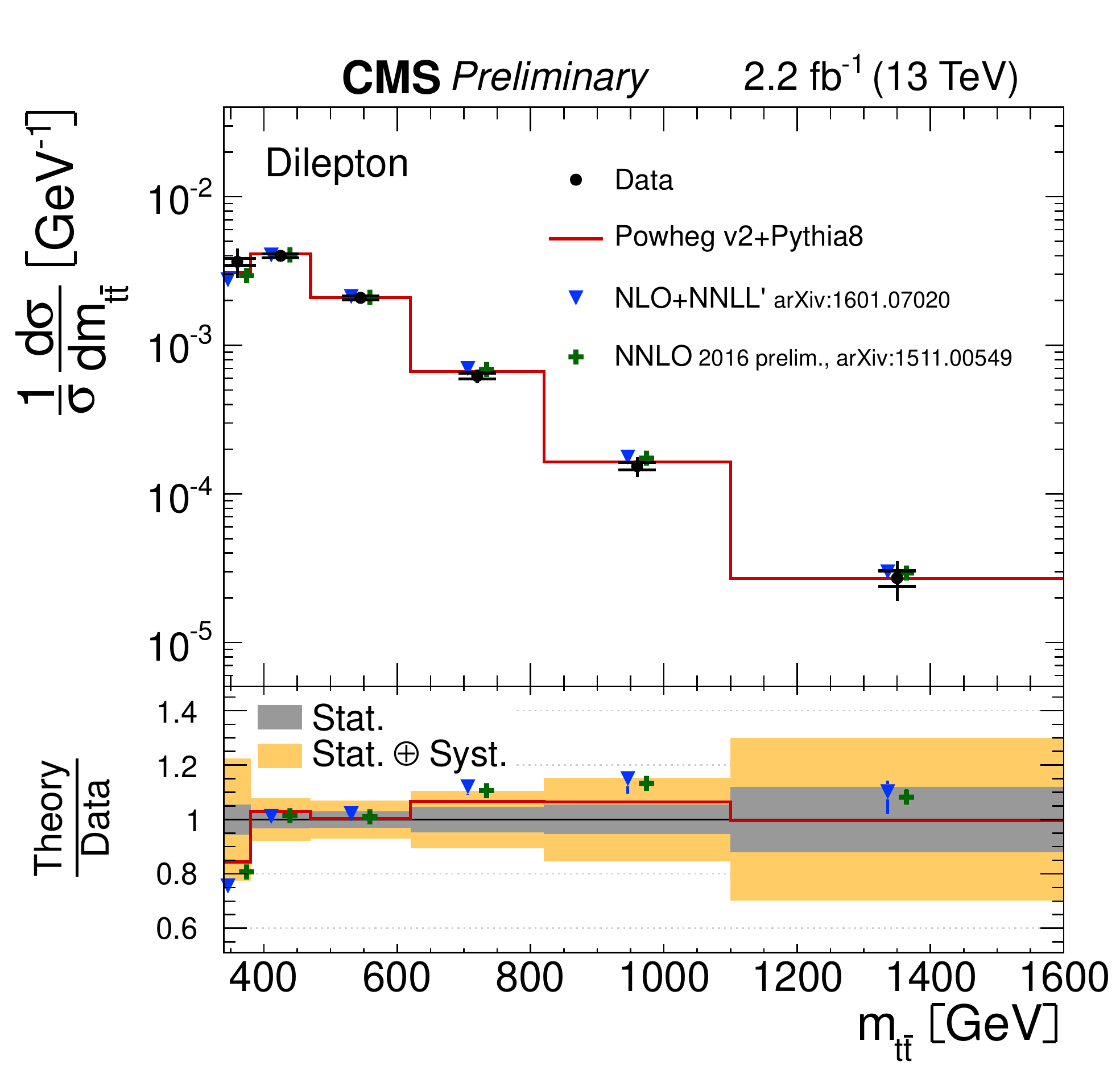}}
\end{minipage}
\caption[]{
  Normalised differential $\ttbar$ production cross section as
  function of the $\pt$ of the top quark or antiquark (left) or the
  invariant mass of the $\ttbar$ system (right). 
  The inner
  (outer) error bar indicate the statistical (total) uncertainty. The
  13 TeV data in the dilepton channel are compared to predictions from
  \textsc{POWHEG}+\textsc{PYTHIA} 8 and to beyond-NLO QCD calculations~\cite{CMS-PAS-TOP-16-011}.
}
\label{fig3}
\end{figure}

The $\ttbar$ kinematics have also been probed, and while it is found
that the experiments will still benefit from higher statistics,
the uncertainties related to the modeling of the parton
shower and hadronization start to be relevant. 
Variables such as the rapidity of the top or the $\ttbar$ system
and the $\pt$ of the $\ttbar$ system show good agreement with most
predictions~\cite{CMS-PAS-TOP-16-011}.
The top quark $\pt$ shown in Fig.~\ref{fig3} (left)
is however observed to be softer in data with respect to the 
prediction from \textsc{POWHEG}+\textsc{PYTHIA 8}.
The comparison to higher order computations in QCD shows a better
agreement up to 300 GeV. At higher $\pt$, the analysis~\cite{CMS-PAS-TOP-16-011},
which is is performed assuming resolved topologies, starts loosing
efficiency due to the merging of the jets, still it results in a
softer top $\pt$ spectrum, with respect to the one predicted by all the
latest calculations.
The latest analyses performed at higher $\pt$ values, in the boosted
regime,
have been performed still using 8 TeV data, but show similar behaviour 
in the slope and normalisation of the top $\pt$,
spectrum up to 1.2 TeV~\cite{Aad:2015hna,CMS-PAS-TOP-14-012}.
These analyses make use of larger jet cones containing the decay
products of each top quark decay.
The measurements in the fiducial and full phase spaces show that the
MC over-predicts the rates by 10-50\% the cross sections at
increasingly high $\pt$. However the predictions suffer from large
parton-shower modeling uncertainties.
A large gain in cross section ($\approx$8) is expected at 13 TeV with
respect to 8 TeV for $M_{\ttbar}>$1 TeV. This will further open the
door to study the properties of boosted top quarks.
It will be interesting to keep probing more precisely, 
the nature of these differences found in data, with respect to the theory predictions.

A particularly interesting variable, as already alluded before,
is the invariant mass of the $\ttbar$ system ($M_{\ttbar}$).
The latest measurements of show a fair agreement in both rate and shape.
The normalised $M_{\ttbar}$ spectrum, showin in Fig.~\ref{fig3} (right),
is measured up to 1.6 TeV in the dilepton channel~\cite{CMS-PAS-TOP-16-011}, 
with an uncertainty which ranges between 5-20\%. The main source of
systematic uncertainty is due to the ambiguity in the corrections
applied to the data to bring it to parton-level. These corrections are
used in unfolding the data and besides taking into account the
experimental resolutions, they have to correct for the acceptance to
the full phase space. As such hadronization, parton
shower scale and matrix-element to parton shower matching related
uncertainties become non negligible as they predict significantly
different evolutions and effects from parton level to final state
particles. These uncertainties seem however to be intrinsic, when
comparing to fixed order calculations which do not include the top
quark decay. 
Further progress towards more precise measurements of this
and other differential distributions passes through improving the MCs
using alternative measurements such as the ones identified in
Section~\ref{subsec:incxsec},
and using particle level definitions.
An example of the latter can be found in the latest measurements of the
$\ttbar$ differential cross section in the $\ell$+jets final state at
13 TeV, released just after this conference, by the CMS Collaboration~\cite{CMS-PAS-TOP-16-008}.

\section{Electroweak production of single top quarks}
\label{sec:ewkprod}

In general, single top production cross sections increase slower with $\sqrt{s}$
than those of $\ttbar$, as the former are quark-initiated
processes and involve electroweak vertices.
An increase by a factor of 2.6 (1.9) at 13 TeV with respect to 8 TeV,
is expected for the {\it t}- ({\it s}-) channel productions~\cite{Kant:2014oha}.
In general good agreement between the current measurements and the
different channels is found at different $\sqrt{s}$ as could be seen
already from Fig.~\ref{fig1} (right). 

\subsection{{\it t}-channel production}
\label{subsec:tch}

Single top quark production via a {\it t}-channel exchange of a W boson is
firmly established at 13
TeV~\cite{CMS-PAS-TOP-15-004,ATLAS-CONF-2015-079}. 
The main characteristic of this channel is the large separation in
pseudo-rapidity between the top quark and the recoiling light quark
jet which is expected to be scattered along the forward region.
The initial 42 $\pbinv$ of 2015 data were sufficient to find evidence for
this process at a 3.5$\sigma$ significance level.
Events with one muon ($\pt>$22 GeV and $|\eta|<$2.1), one $b$-tagged
jet and one forward jet (both with $\pt>$40 GeV and
$|\eta|<$4.7)~\cite{CMS-PAS-TOP-15-004}. An extra requirement on the
transverse mass of the lepton and missing transverse energy ($M_{\rm
  T}>$50 GeV) is used to reduce the contamination from QCD multijets.
The distribution of the forward jet in pseudo-rapidity was fit to
measure the {\it t}-channel cross section with a statistics dominated uncertainty of 42\%.
With the full 2015 integrated luminosity the ATLAS collaboration has 
made use of a multivariate discriminator to
explore the main characteristics of the {\it t}-channel process: the
forward quark jet, top mass related variables and the W boson Jacobian
peak~\cite{ATLAS-CONF-2015-079}. 
The backgrounds are predicted from simulation but validated in control regions
dominated either by $\ttbar$ (2 b-tags) or W+jets (1 loose b-tag).
A fit to the multivariate discriminator is used to extract the signal
strength with a 20\% uncertainty, dominated by signal
modeling-related uncertainties.
Figure~\ref{fig4} (left) shows the distribution of the variable used
in the fit in events with a positively charged muon.
Soon after this talk has been given the CMS Collaboration has also
updated the analysis with the full 2015 dataset, and using a similar approach,
fitting a multivariate discriminator, and attaining a 15\% total uncertainty~\cite{CMS-PAS-TOP-16-003}.
The production rate for the top or the antitop-quark,
expected to differ due to the PDF composition of the proton,
is measured separately in both analysis quoted before.
The result is verified to be in agreement with the
predictions ($\sigma(tq)/\sigma(\bar{t}q)\approx$ 1.68)
but still suffers from large uncertainty.
With increased integrated luminosity and improved analysis this is
expected to provide a further handle on the proton PDF.

\subsection{{\it s}-channel production}
\label{subsec:sch}

The resonant production of single top quarks ({\it s}-channel)
is a process which it  difficult to discriminate at the LHC.
The final state comprises the decay products of a top quark and an
extra $b$ jet. This
 can be easily be mimicked by $\ttbar$, {\it t}-channel and
W+heavy flavour productions.
In addition the cross section is expected to be very low,
$\mathcal{O}(10)$ pb, when compared to that of the processes referred before.
However given the unique properties of a {\it s}-channel, 
there is interest in measuring accurately this process since it
tests directly the CKM matrix element $V_{\rm tb}$ and it offers the
potential to seek for BSM physics such as the production of a charged
Higgs boson.
After analysing the full 7+8 TeV datasets, CMS could set a
$4.7\sigma_{SM}$ upper limit at 95\% CL, on the production cross section of
tops through the {\it s}-channel, being the expectation
$3.1\sigma_{SM}$ at 95\% CL~\cite{Khachatryan:2016ewo}.
The analysis, based on a multivariate discriminator, was found to
suffer from theory uncertainties related to the modeling of $\ttbar$
but also the {\it t}- and {\it s-} channels.
Using the 8 TeV dataset, the ATLAS collaboration has employed a matrix element technique
to discriminate against the main backgrounds and found evidence for 
this process at 3.2$\sigma$ level in fair agreement with the 3.9$\sigma$
expectation. The extracted cross section is observed to be in agreement with
the SM prediction within the 37\% uncertainty attained.
The uncertainty on the modeling of  $\ttbar$,{\it t}- and {\it s}-channels 
contribute to 13\% of the uncertainty, while experimental uncertainties
such as jet energy scale and b-tagging contribute at the level of
15\%. The statistics of the simulation is also verified to be a
limiting factor of the analysis, at the level of 12\%.

Prospects for Run 2 of the LHC are not as bright as for the other
top-related processes. The {\it s}-channel cross section
increases only by a factor of 1.9 between 8 TeV and 13 TeV,
in contrast with the increases expected for $\ttbar$ and the {\em t}-channel.
Nevertheless this remains an interesting process to study in more
detail in the future.

\subsection{$V_{\rm tb}$ measurements from production rates}
\label{subsec:vtb}

\begin{figure}
\hfill
\begin{minipage}{0.44\linewidth}
\centerline{\includegraphics[width=1.0\linewidth]{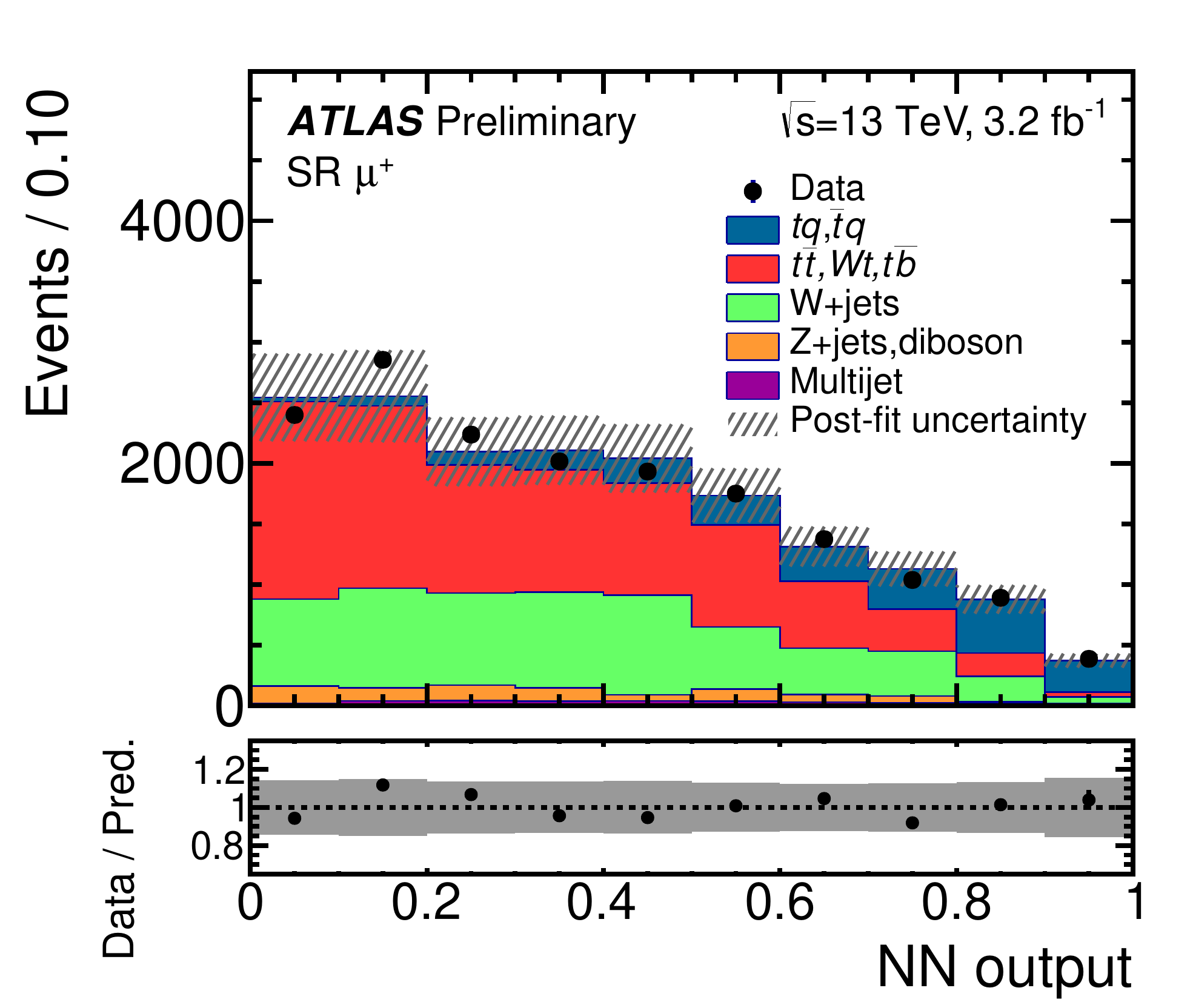}}
\centerline{\includegraphics[width=1.0\linewidth]{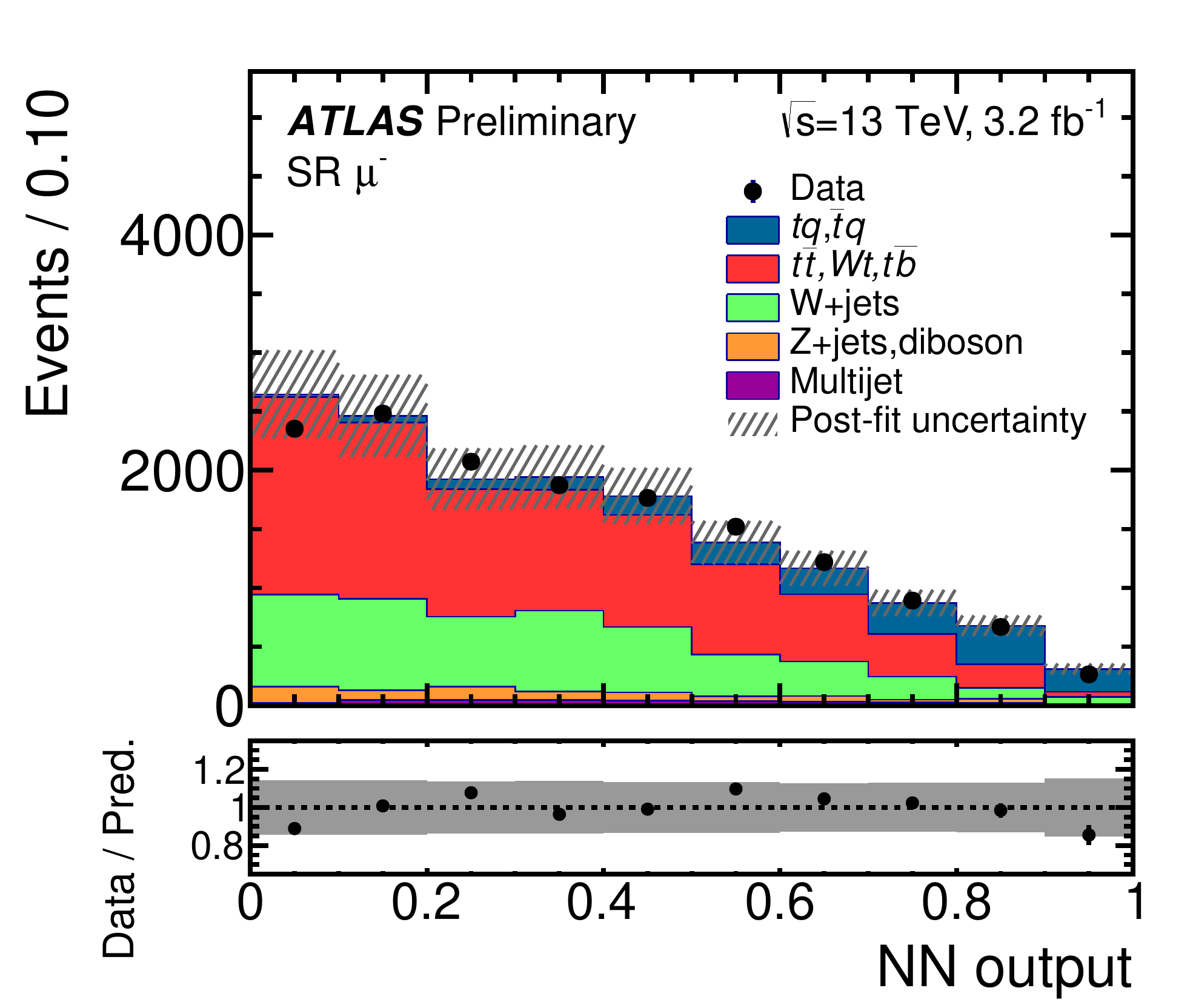}}
\end{minipage}
\hfill
\begin{minipage}{0.55\linewidth}
\centerline{\includegraphics[width=1.0\linewidth]{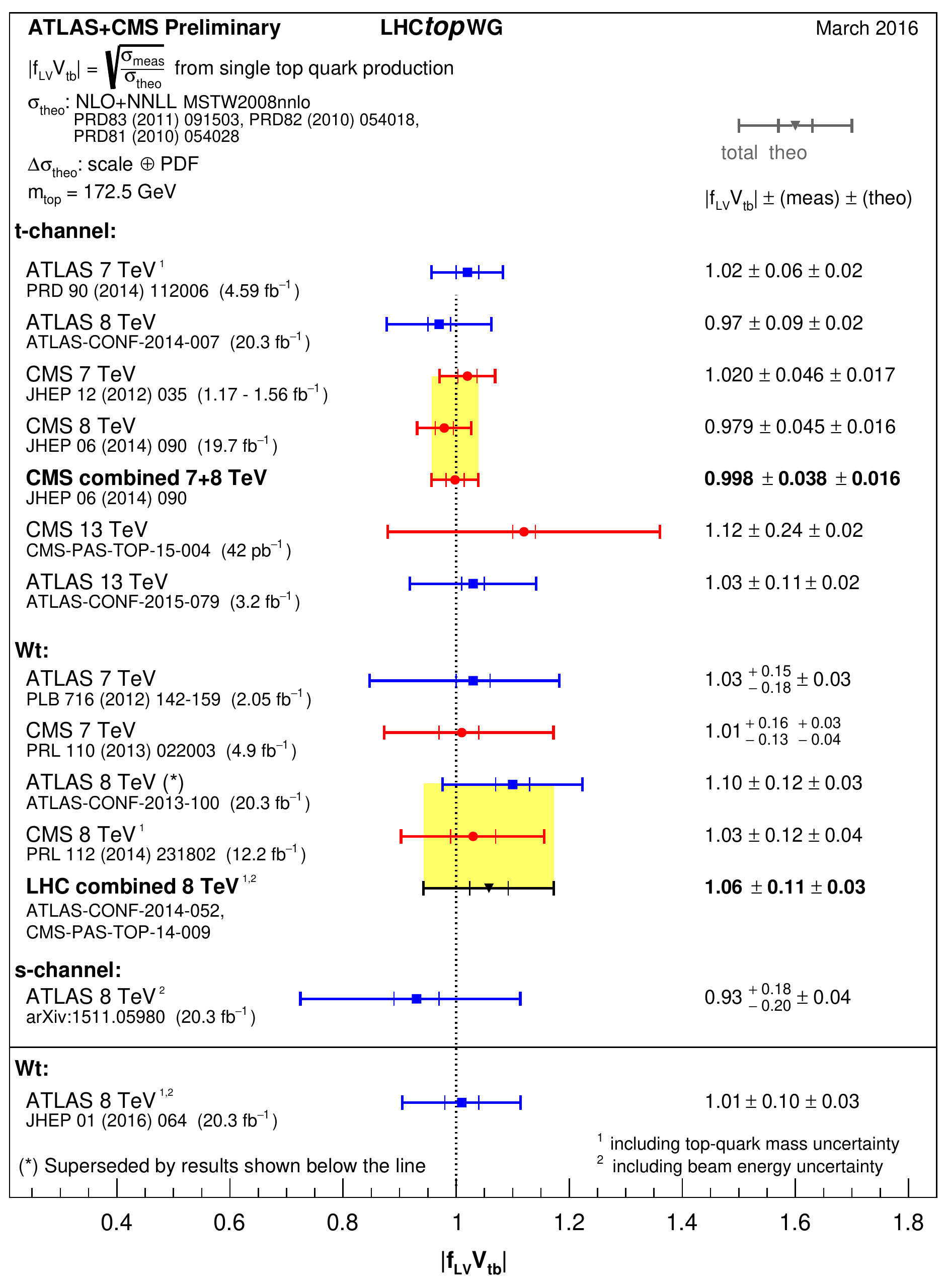}}
\end{minipage}
\caption[]{
 Left: Multivariate discriminator distribution for the $\mu^+$ (top)
 and $\mu^-$ (bottom) channels
 in the {\it t}-channel analysis at 13 TeV~\cite{ATLAS-CONF-2015-079}. Signal
 and backgrounds are normalised to the result of the fit being the
 post-fit uncertainty represented as a hatched band.
 Right: Summary of the extractions of the CKM matrix element $V_{\rm
   tb}$ from different single top measurements performed at 7 and 8 TeV.
 The contribution from the theory uncertainty to the final value is represented by the inner error bar.
}
\label{fig4}
\end{figure}

The cross section for the electroweak production of top quarks is
sensitive to $|V_{\rm tb}|^2$,
and as such can be explored to extract this parameter. The most
precise measurement attains 4\% uncertainty in the measurement of
$|V_{\rm tb}|$, as shown in Fig.~\ref{fig4} (right). 
The uncertainty is equally shared between experiment and theory as
$\Delta V_{\rm tb}/V_{\rm tb}\approx \frac{1}{2}(\Delta \sigma_{\rm exp}/\sigma_{exp}\oplus \Delta \sigma_{\rm th}/\sigma_{\rm th})$,
where $\sigma_{\rm exp}$ ($\sigma_{\rm th}$) is the experimental
measurement (theory prediction) for the production cross section.
Prospects for Run 2 may include more precise theory predictions, now
know at NNLO QCD for the {\it t}-channel production~\cite{Brucherseifer:2014ama},
and refined techniques in the extraction of the cross section,
including a more detailed evaluation of the effects on the decay if $|V_{\rm tb}|<1$.
One should also bear in mind, that these results can be compared with
a direct measurement of $B(t\rightarrow Wb)$ in $\ttbar$ events.
Under the assumption of CKM unitarity and no additional sequential
generation of quarks, the latter yields currently a measurement of $|V_{\rm tb}|$ with a 1.6\% uncertainty~\cite{Khachatryan:2014nda}. 
A combination of the single top quark
cross section in the {\it t}-channel, with the measurement of
$B(t\rightarrow Wb)$ allows to indirectly
extract the top quark width with an uncertainty of  10.5\%.
In Run 2 it will be important to improve on the current
knowledge of the top quark width by performing also direct
measurements of this quantity. Further news on the latest top quark properties
measurements at the LHC have been reported in this conference by E. Monnier~\cite{monniere}.

\section{Conclusions}
\label{sec:conclusions}

The ATLAS and CMS collaborations are finalising the analysis of the
first 13 TeV data collected in 2015. These data can be viewed as a drop in the
ocean ahead of us. While the recent measurements of the cross sections
(inclusive and differential) is starting to compete with the theory
predictions in terms of uncertainty,
they are approaching a systematics-limited regime.
To overcome this limitation in the long run, alternative measurements
and tunes of our simulations are needed.
Some have been identified along this writeup. Any extraction of fundamental parameters such as
the pole mass, $V_{\rm tb}$, or others, is expected to be affected by extrapolation
uncertainties from the fiducial region to the full phase space.
These include non-perturbative effects which are modeled by the MC simulations
and can at most be tuned using control regions or ``peripheral''
measurements of the radiation environment and underlying event in top
quark events.
Carrying these measurements to improve the modeling of the top
signals is crucial, towards probing optimally the properties of this
quark with the full data expected to be acquired in Run 2 of the LHC.

\section*{Acknowledgments}

The organizers of the conference are acknowledged for the invitation
and for the warm environment which has hosted interesting discussions.
The CMS and ATLAS Top group conveners, Andrea Giammanco, Fr\'ed\'eric D\'eliot  and Mark Owen, 
are also thanked for helping review the talk and these proceedings.

\end{document}